\documentclass[12pt]{iopart}
\usepackage{verbatim}

\usepackage{graphicx}
\usepackage{times,cite,w-thm}

\begin{document}

\title{E$\times$B analyzer for fast ion temperature measurements in scrape-off layer of the ASDEX Upgrade and COMPASS tokamaks}

\author{M. Komm$^{1}$, M. Ko\v{c}an$^2$,D. Carralero$^3$ J. St\"{o}ckel$^1$, A. Herrmann$^3$ and R. P\'{a}nek$^1$}
\address{$^1$ Institute of Plasma Physics AS CR, v.v.i., Za Slovankou 3, 182 00 Prague 8, Czech Republic}
\address{$^2$ ITER Organization, Route de Vinon sur Verdon, 13115 St Paul Lez Durance, France  }
\address{$^3$ Max-Planck-Institut f\"{u}r Plasmaphysik, EURATOM Association, Boltzmannstr. 2, 85748 Garching, Germany}

\ead{komm@ipp.cas.cz}

\begin{abstract}
Ion temperature in the scrape-off layer is an important quantity but rarely measured at high temporal resolution. 
In order to achieve fast measurements during intermittent events such as blobs, a novel E$\times$B analyzer has been designed. 
The measurement with the E$\times$B analyzer does not require voltage sweeping but records ion current on DC biased collectors, meaning that the temporal resolution is only limited by the sampling frequency of the data acquisition system and electronics of the amplifiers, which exceeds the typical timescale of events of interest ($\sim$10 $\mu$s).

The analyzer is equipped with entrance slit of optimized knife-edge shape, which reduces selective ion losses and improves overall ion transmission. The design of the analyzer is presented together with first acquired signals and reconstructed temperatures.
\end{abstract}

\submitto{\PPCF}
\maketitle

\section{Introduction}
The ion temperature measurements in the scrape-off layer (SOL) are of high interest because of incomplete understanding of transport processes and plasma-wall interaction. Although models predicting such transport exist \cite{esel}, the ion temperature and its fluctuations are neglected in most of the models partially due to the lack of systematic experimental data with temporal resolution on the turbulent and blob filament time scale. Such cold ion approximation is in contradiction with measured profiles of $T_i$ in SOL, which typically exceed those of $T_e$ \cite{kocan-ti-profile}. Recent modelling \cite{esel-recent}\cite{esel-recent2} with finite $T_i$ showed that ion temperature influences drift wave growth rate and filament propagation speed. However, the role of $T_i$ fluctuations is not yet completely understood.

The measurements of ion temperature cannot be obtained using the standard sweeping Langmuir probe because of the excessive electron current, which masks the changes of ion current for probe potential above the plasma potential. Several techniques for for measuring the ion temperature in the tokamak scrape-off layer have been developed, the most commonly used being the Katsumata probe \cite{katsumata} and the retarding field analyzer (RFA) \cite{rfa}\cite{brunner}\cite{pitts}. Although it is possible to achieve ion temperature on the filament timescale ($\mu$s) using such diagnostics \cite{kocan-ion-fluct}, it is technologically challenging and requires condition averaging.

In order to facilitate fast $T_i$ measurements, an E$\times$B analyzer has been used earlier on DITE \cite{matthews} and Asdex\cite{staib} tokamak. We have refined the design of the ExB analyzer with aid of numerical simulations. 

\section{Principle of the E$\times$B analyzer}
\label{intro}

\begin{figure}[htbp]
\begin{center}
\includegraphics[scale=0.66]{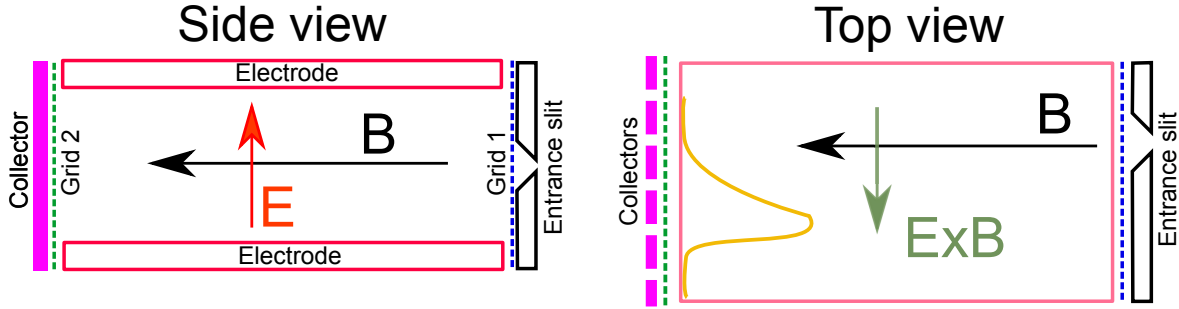}
\caption{Schematic top and side view of the E$\times$B analyzer.}
\label{schema}
\end{center}
\end{figure}

The E$\times$B analyzer consists of an array of collectors and a pair of planar electrodes located in a cavity behind a slit plate perpendicular to the tokamak magnetic field B (see Fig. \ref{schema}). The slit plate protects the internal components from the plasma heat fluxes and reflects the electrons back into the plasma. The dispersion of the ion guiding centers from the slit axis parallel to B (i.e. along the collector array) $\Delta_x$ can be expressed as:
\begin{equation}
              \Delta_x = \frac{E}{B}L \frac{1}{v_{||}},
\label{def}
\end{equation}
where $L$ is the length of the electrodes inside cavity and $v_{||}$ is the ion parallel velocity. \cite{matthews}.  The electric field $E$ is created by a potential difference between the electrodes (dubbed top and bottom), which are separated by distance $d$.
\begin{equation}
             E = (V_{top}  - V_{bottom})/d
\label{def-E}
\end{equation}

The parallel velocity consists of the original velocity at which ion enters the sheath in front of slit plate and an additional component due to acceleration caused by potential inside the cavity $V_{mid}$
\begin{equation}
            v_{||} = v_{i} + \sqrt{\frac{2e (-V_{mid} + V_{plasma})}{m_i}}.
\label{eq-acc}
\end{equation}
, where  $V_{mid}$ is the potential along ion beam path inside the cavity, which is typically located half way between the planar electrodes .
\begin{equation}
            V_{mid} = (V_{top}  + V_{bottom})/2
\label{eq-vmid}
\end{equation}
 Since in the experiment the magnitude of potential $V_{mid}$ is typically much larger than the plasma potential, we neglect the later and assume it equal to zero. The acceleration inside cavity to $V_{mid}$ introduces maximum deflection $ \Delta_{x,max}$ for particles with zero initial velocity.
\begin{equation}
           \Delta_{x,max} =\frac{(V_{top} - V_{bottom}) L}{B d \sqrt{(e -(V_{top}  + V_{bottom})/m_i))}}
\label{eq-max-dx}
\end{equation}
In the experiment one of the electrodes (in case of AUG $V_{top}$) is typically fixed at zero voltage in order not to exceed, so $\Delta_{x,max}$ depends only on the applied voltage on $V_{bottom}$.
\begin{equation}
           \Delta_{x,max} =\frac{L \sqrt{-V_{bottom}}}{B d \sqrt{e/m_i}}
\label{eq-max-dx-vbot}
\end{equation}
The choice of $V_{bottom}$ hence determines the span over which the ions will be spread at the collectors.

Measurements by RFAs in the tokamak SOL showed that ions in the high-energy range of the parallel ion distribution are characterized by the Maxwellian distribution of the parallel speeds \cite{kocan-jnm} and therefore suitable for fitting by using the formula:
\begin{equation}
            I_c \sim exp \left ( - {v^2_{i}} \over {T_i} \right) 
\end{equation}
or by replacing the parallel velocity by displacement $ \Delta_x$ 
\begin{equation}
I_c(\Delta_x)=I_0 exp \left (  {E^2 L^2} \over {B^2 \Delta_x^2 T_i} \right). 
\end{equation}
In practice we measure currents from collectors at different $\Delta_x$ and by fitting $I_c(\Delta_x)$ with an exponential, we can obtain $T_i$.

%\begin{equation}
 %             \delta_x   =  T_{L} v_D = \frac{2 \pi m_i E}{q B^2},
%\end{equation}

\section{Design considerations}
\subsection{Probe head dimensions}
The probe head is designed for use on the horizontal reciprocating manipulators on AUG and COMPASS, which have compatible interfaces. The reciprocation allows to acquire measurement inside the SOL without damaging the diagnostics by enormous heat fluxes. On AUG,  multiple strokes during a discharge with typical stroking frequency 1 Hz are possible, COMPASS manipulator allows single reciprocation due to limited duration of the discharge.  The manipulator imposes upper limit on the diameter of the probe head  - maximum 60 mm is allowed. Due to high heat fluxes in the SOL especially during ELMs, the analyzer has to be protected by a carbon cover, which leaves effectively some 50 mm of inner space for the analyzer itself. This limit determines the maximum size of the inner cavity and as such is crucial for the design of the diagnostics. The remaining parts of this section describe parts of the analyzer which are of special attention.

\subsection{Entrance slit - geometry}
\label{section-slit}
The entrance slit is the plasma-facing part of the analyzer and serves to stop electrons from entering the cavity. Unlike the RFA, where parasitic electron current itself degrades the measurement, in case of the analyzer, it has to be avoided mainly due to unwanted side effects. The electrons can ionize neutrals within the analyzer cavity, which would then contribute to the collector currents. The main goal of the design thus to optimize  the slit geometry in order to allow measurable ion current inside the cavity (while acting as a barrier for electrons). At the same time the slit has to be capable of withstanding the high heat loads when operating in the vicinity of last closed flux surface or during the ELMs.
\begin{figure}[htbp]
\begin{center}
\includegraphics[scale=0.33]{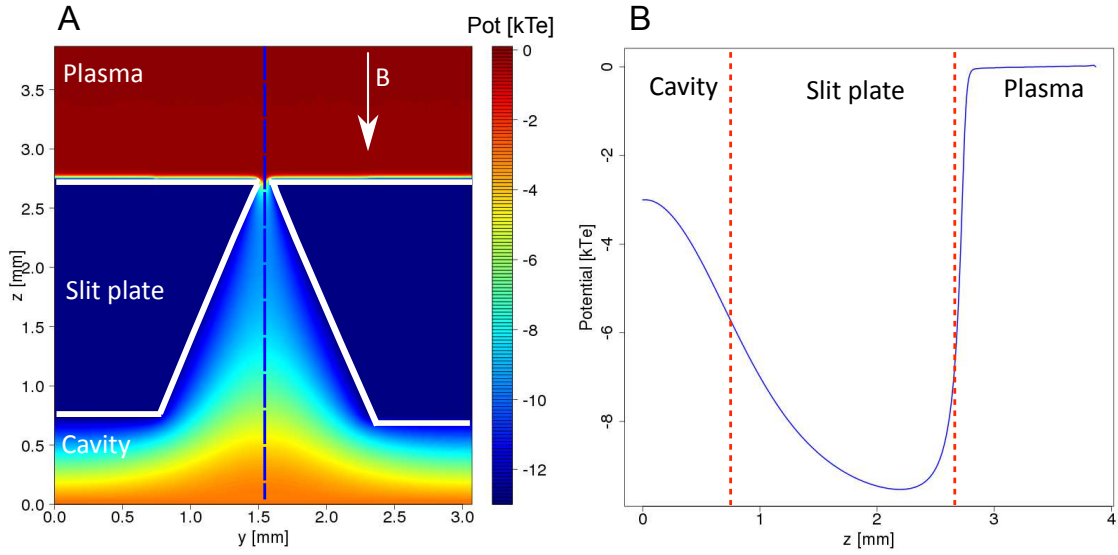}
\caption{Potential distribution around the entrance slit (left) and profile along the slit axis (right). Slit width (100 $\mu$A) equals to $13 \times \lambda_D$.}
\label{spice}
\end{center}
\end{figure}

In previous experiments with RFAs (which also required biased slit)[REF - Pitts thesis] \cite{kocan-jnm}, the slit width was chosen to be comparable to $\lambda_D$ to prevent potential leaking inside the cavity (which would admit electrons to surpass the slit). However, dedicated particle-in-cell (PIC) simulations using a SPICE2 code\cite{komm-textor}\cite{renaud-spice} revealed that this criterion can be somehow relaxed and slits as wide as $10\times\lambda_D$ still do not admit electrons inside the cavity, as shown in Fig. \ref{spice}. The simulation was performed for the worst case scenario, which can be foreseen on AUG ($n_e$ = 1$\times$10$^{19}$ m$^{-3}$, $T_e$=10 eV, $T_i$ = 20 eV, B= 1.9 T) with very short Debye length. The figure shows the distribution of self-consistently calculated potential (A) as well as the potential profile along the slit axis (B). The profile illustrates two important features of the simulated slit. The potential barrier for the electrons is as deep as -9 k$T_e$ (the slit is biased to -13 k$T_e$, meaning 260 V below the plasma potential). At the same time the space charge due to ion beam does not result in potential structures surpassing the plasma potential, so the space charge does not block the ions from entering the cavity.  A thinner slit ($~\sim \lambda_D$) would ensure full potential barrier at -13 $kT_e$ but it would also admit much smaller (and thus more difficult to measure) ion current inside the cavity.

Increasing the slit length is a simple mean to increase the collector currents. Since all the particles inside the cavity should be affected by the same potential, the slit has to oriented so that the longer side is parallel to the collector array (ie. perpendicular to the individual collectors).  For this reason, the slit should not be larger than the collector width ($\sim$1 mm). 

\begin{figure}[htbp]
\begin{center}
\includegraphics[scale=0.5]{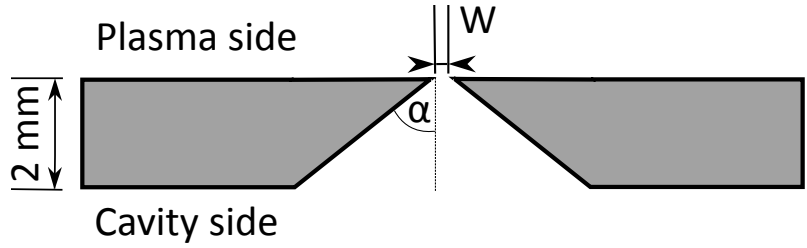}
\caption{Schematic view of the entrance slit.}
\label{schema-slit}
\end{center}
\end{figure}

\begin{figure}[htbp]
\begin{center}
\includegraphics[scale=0.1]{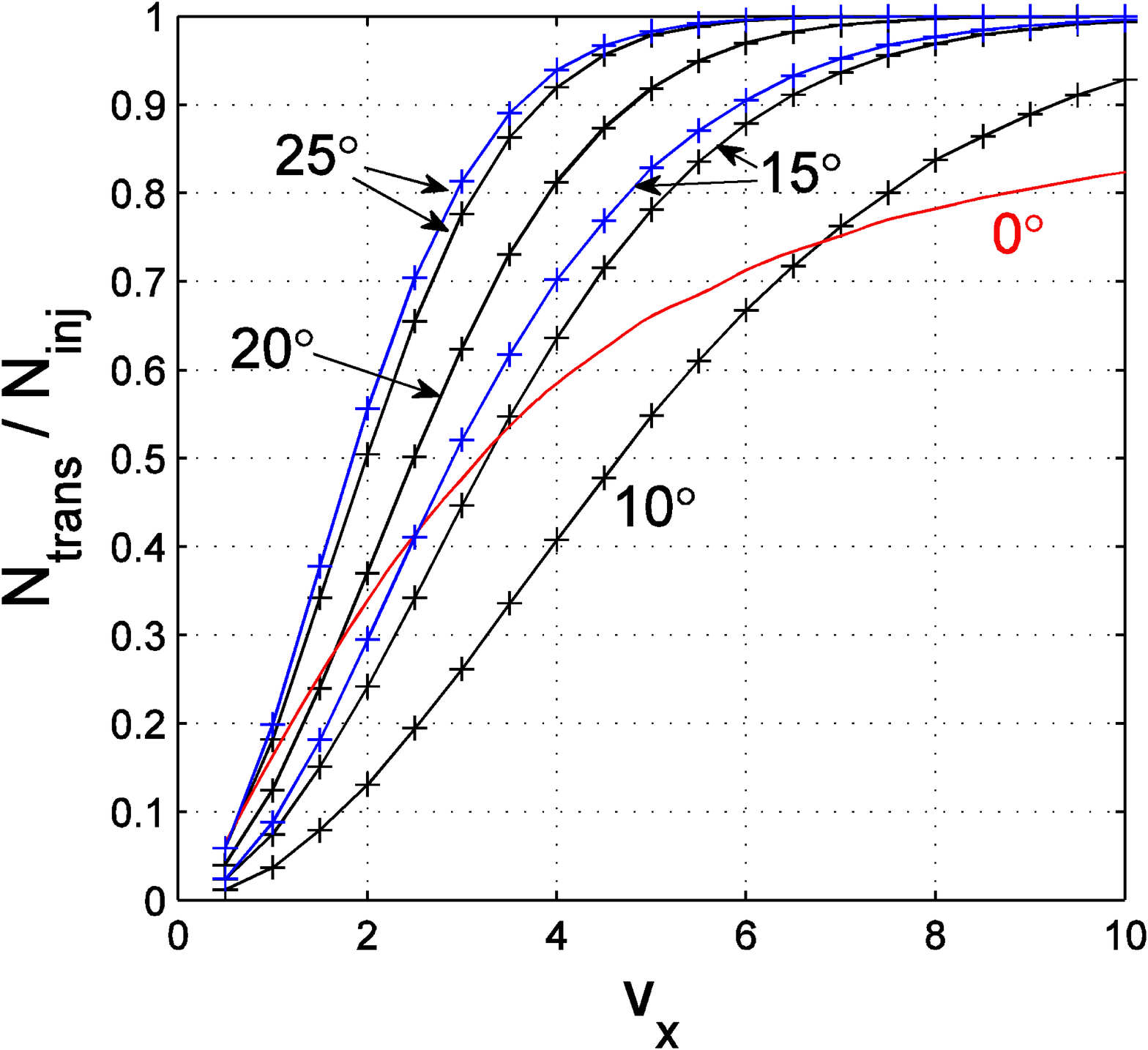}
\caption{Transmission factor $N_{trans}/N_{inj}$ as a function of parallel velocity $v_x$ for varying angle $\alpha$ of the slit (2 mm thick). The red line shows the transmission factor of a rectangular slit. }
\label{trans}
\end{center}
\end{figure}

Another mean of increasing the admitted current without enlarging the slit is to increase the ion transmission by using a knife-edge shaped slit \cite{pitts} (see Fig. \ref{schema-slit}) instead of a rectangular one. The transmission was first simulated using a Monte-Carlo code for varying angle $\alpha$ of the opening between 0 and 25 degrees. Deuterons with mono-energetic parallel speeds and Maxwellian perpendicular velocities (characterized by $T_i$=40eV) were launched at random position along the slit entrance. Simulated slit was  100 $\mu m$ wide and $1000 \mu m$ long, cut into 2 mm thick slit plate. The slit transmission factor (i.e. the number of transmitted ions $N_{trans}$ to the total number of injected ions $N_ {inj}$) calculated for different $\alpha$ and $v_{||}$ is shown in  Fig. \ref{trans}. The ion transmission increases with $\alpha$ and is almost insensitive to $v_{||}$ for large $v_{||}$ from which $T_i$ is derived in experiment. This is a substantial improvement over the rectangular slit (shown as red line in the figure). It should be noted that all ions are accelerated in the parallel direction by 2-10 $T_e$ due to the potential drop in the Debye sheath in front of the slit plate and thus arrive at the slit plate with relatively large $v_{||}$. The angle $\alpha=20^\circ$ was chosen for the ExB analyzer. 

\begin{figure}[htbp]
\begin{center}
\includegraphics[scale=0.4]{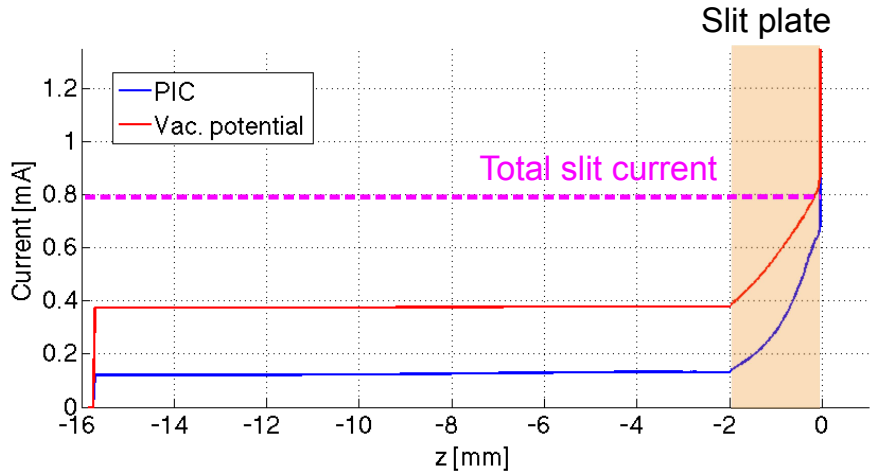}
\caption{Current flowing through the analyzer cavity for vacuum potential (red) and self-consistent potential calculated by SPICE3 code (blue). Theoretical current based on current density arriving at the slit plate and slit cross-section marked in magenta.}
\label{slit-transmission}
\end{center}
\end{figure}

It has to be noted that such high transmission was not achieved in experiment. When comparing the input current (calculated by the slit plate current density integrated over the slit cross-section) with total ion current registered by all collectors, the effective transmission was typically close to $\sim 15\%$, which is similar to RFA experiments \cite{brunner}\cite{kocan-jnm}. The strong attenuation is caused by electric fields inside the slit volume, which deform the ion trajectories and in many cases prevent them from passing through the slit. The problem was simulated using a 3D particle-in-cell code SPICE3\cite{komm-katsumata} and the effective slit transmission is shown in Fig. \ref{slit-transmission}. Despite the absence of siginificant space charge effects inside the slit volume, the slit transmission for self-consisted potential profile was only $20\%$, while for vacuum potential it has reached $50\%$ (which is however still lower than MC code prediction). 

\begin{figure}[htbp]
\begin{center}
\includegraphics[scale=0.20]{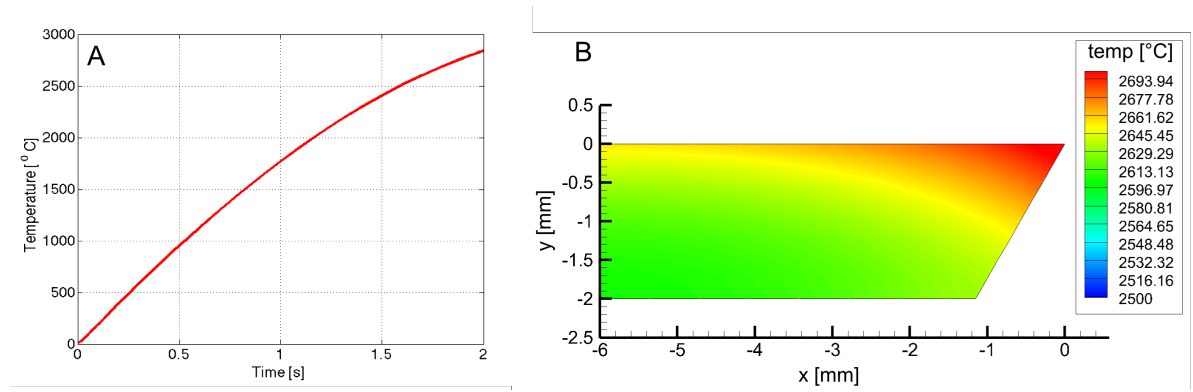}
\caption{Temperature of the hottest point at the slit during a 2 s pulse with 10 MW/m$^2$ load.}
\label{FEM}
\end{center}
\end{figure}

In order to study the power handling of the slit plate under the plasma exposure, the slit plate was modelled by the two-dimensional finite difference heat transfer code which accounts for the variation of tungsten properties with the temperature. In the simulation the plasma facing part of the slit was irradiated by the steady-state heat flux of 10 MW/m$^2$, expected during the ELM in the AUG far SOL, lasting 2 seconds. Such long heat pulse duration corresponds to rather extreme conditions. The typical ELM duration and frequency in AUG are, respectively, $\sim$4ms and 50Hz, meaning that the slit plate will on average experience the heat loads of about 2MW/m$^2$ during the ELMy H-mode. Moreover, the probe is typically maintained in the plasma only for several tens of miliseconds.
The heat transfer from the slit to the support structure was neglected. Fig \ref{FEM}A shows the time evolution of the temperature of the slit leading edge (at which the highest temperature was observed), with $T_{max}$=2900K, well below the tungsten melting temperature (3695K). Fig \ref{FEM}B shows the temperature distribution within the slit (for t=1.7 s), which shows very small difference of temperatures. Despite the exaggerated pulse duration, the melting of the slit leading edges is therefore highly unlikely.

\subsection{Space charge effects inside the cavity}
\begin{figure}[htbp]
\begin{center}
\includegraphics[scale=0.3]{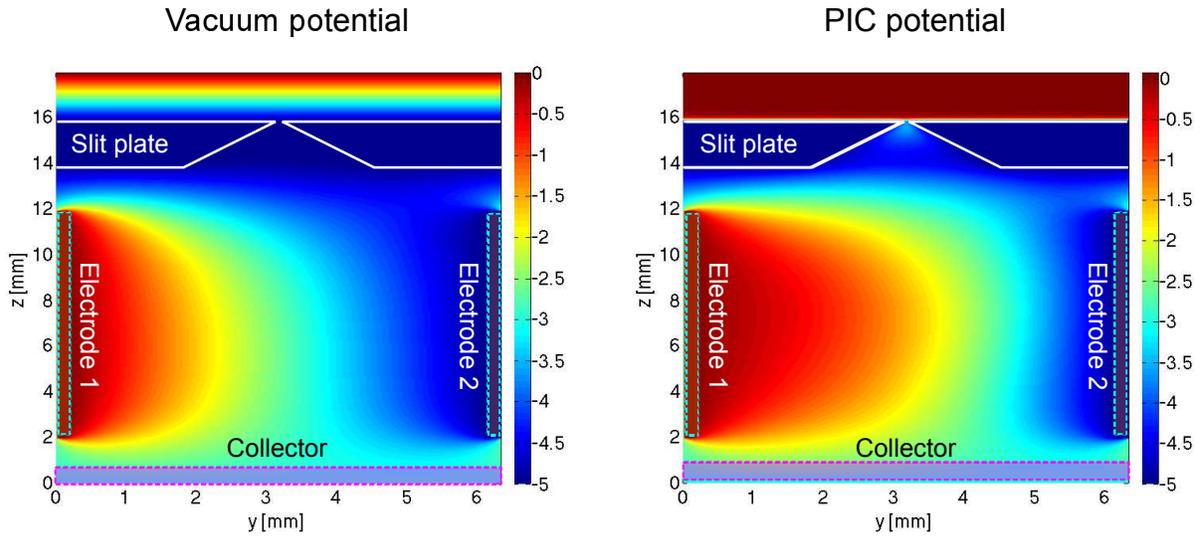}
\caption{Potential calculated by SPICE3 code for a slit 0.1x1.0 mm. Vacuum potential (left) and self-consistent potential (right).}
\label{potential-3d}
\end{center}
\end{figure}

In the previous section, the slit was optimized to admit maximum ion current inside the cavity. Since the cavity is substantially larger (33x6x15 mm$^3$) than in case of modern RFAs , concerns about the possible build-up of space charge inside the cavity had to be addressed. Such space charge would be deprimental for the measurement for a number of reasons: It could prevent the ions from reaching the collectors, deform their velocity distribution function and suppress the E field created by the electrodes. Analytical calculations following simple analytical approach \cite{brunner} indeed indicated that for the expected ion fluxes the space charge would be well developed.  The problem was first targeted by 2D particle-in-cell simulations, which confirmed potential structures surpassing the plasma potential. However, dedicated 3D particle-in-cell simulations using SPICE3 code \cite{komm-katsumata} showed that the space charge is negligible, as shown in Fig.\ref{potential-3d}   (0.1x1.0 mm slit was simulated in COMPASS SOL plasma $T_e = 20 eV$, $n_e$ = $1 \times 10^{18}$ m$^{-3}$, B=1 T ). This is due to the fact, that simulating slit of finite width (which is not possible in 2D approach) brings additional attenuation of the ion flow on slit sides (the slit length is comparable to $r_{Li}$). At the same time it allows to capture how the ion beam expands inside the cavity volume, which reduces local charge density.

\section{Signal analysis}
Using a guiding center approach, one could expect that the current density distribution on the collectors could be directly associated with the shape of the parallel ion velocity distirbution function. However, since the cavity size is not significantly larger than the ion Larmor radius, finite Larmor effects have to be taken into account. Ions are admitted inside the analyzer cavity through a thin slit, which forces them to follow a cycloid with node separation \cite{matthews}

\begin{equation}
\Delta_n = \frac{2 \pi m_i E_y }{e B^2},
\end{equation}

\begin{figure}[htbp]
\begin{center}
\includegraphics[scale=0.33]{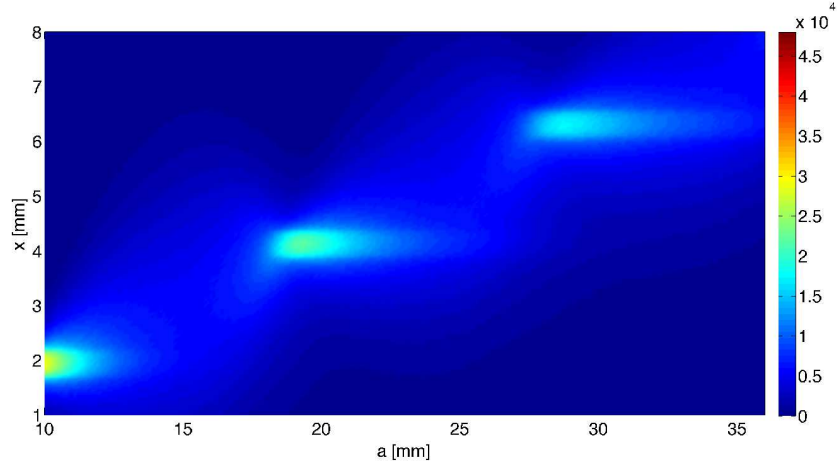}
\caption{Current density (in arbitrary units) in the direction of E$\times$B drift (x) as a function of cavity length $a$. Particles are injected at $x$=0 and $a$ = 0.  }
\label{nodes}
\end{center}
\end{figure}

Depending on the length of the cavity, the current density profile can be significantly modified, as shown in Fig. \ref{nodes}, where current distribution along the $x$ coordinate (direction of the E$\times$B drift) are shown as a function of cavity length $a$.

These profiles were obtained using a simple test particle Monte Carlo (MC) code, which calculates ion trajectories inside the cavity. The input parallel distribution velocity function was taken from a 1D quasineutral kinetic model of the scrape-off-layer \cite{qpic}, which satisfies Bohm criterion. The perpendicular velocity distribution function was assumed Maxwellian.  The particles were generated at random positions inside the slit $z(t_0)$ and their motion was then resolved analytically assuming static homogeneous B field. The time needed to  pass through the cavity is $t_{fin}$.
\begin{equation}
t_{fin} = v_x/L,
\end{equation}
where x direction is parallel to the magnetic field.  The particle reaches its final position $z(t_{fin})$ at x=L
\begin{equation}
z(t_{fin}) = \rho_L (sin \psi - sin(\psi + \Omega_L t_{fin})) + \frac{E_y}{B} t_{fin} + z(t_0),
\end{equation}
Here $\psi$ is the phase between $y$ and $z$ velocity component (generated from uniform distribution), $\rho_L$ is the Larmor radius and $\Omega_L$ the Larmor frequency.

The MC simulation assumes constant drift velocity (and so $E_y$) and potential equal to $V_{mid}$ along the ion beam path. Particle-in-cell simulations in the previous section shown, that self-consistent E field can have significant impact on slit transmission, so the MC code was benchmarked with SPICE3 simulation. Since a full 3D3V simulation of the entire cavity is computationally extremely demanding (especially in terms of required RAM), only a single selected case could be compared with the much easier MC calculations. This simulation was carried out for the same plasma parameters as in section \ref{section-slit} and required a grid of 192$\times$256$\times$1280 cells, the largest case calculated by SPICE3 to date, with complete runtime of approximately 2 months. One electrode was left at 0, the other biased to -100V. This is lower voltage difference than used in experiment but allowed to capture the current density distribution within the available volume (the cavity had to be shortened in the direction of E$\times$B drift).

\begin{figure}[htbp]
\begin{center}
\includegraphics[scale=0.33]{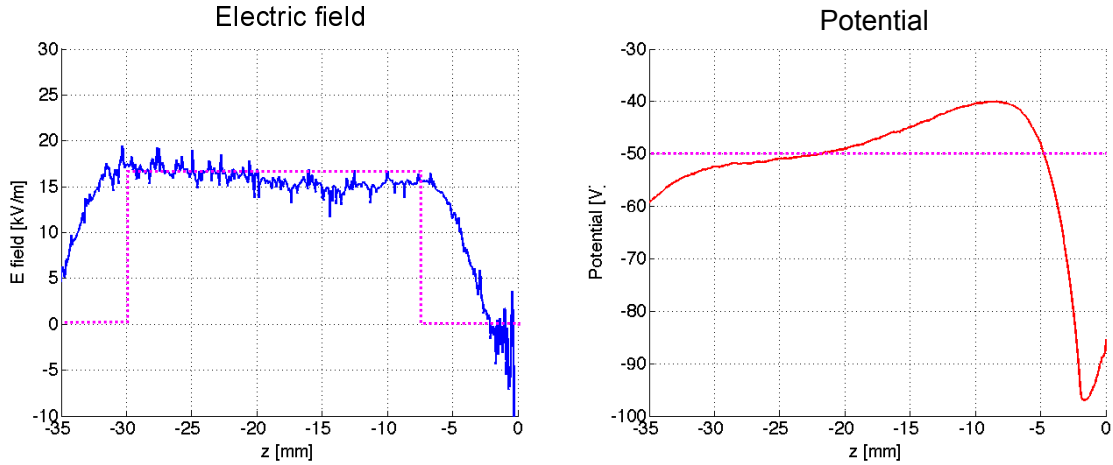}
\caption{Profiles of $y$ component of E field and potential along the B field inside analyzer's cavity. PIC results are plotted in solid lines, MC assumptions dashed. Position of electrode in PIC simulation indicated in orange.}
\label{PIC_vs_MC_pot}
\end{center}
\end{figure}

The profiles of $y$ component of E field and potential along the magnetic field line which intercepts the slit are shown in Fig. \ref{PIC_vs_MC_pot}. It can be seen that MC assumptions (plotted in dashed lines) are simplifying the real profiles from PIC simulation but the difference is not dramatic. In order to get better match in terms of applied electric field, one can shorten the cavity length in the MC simulation, so that the effect on ions will be comparable to PIC run. 

\begin{figure}[htbp]
\begin{center}
\includegraphics[scale=0.33]{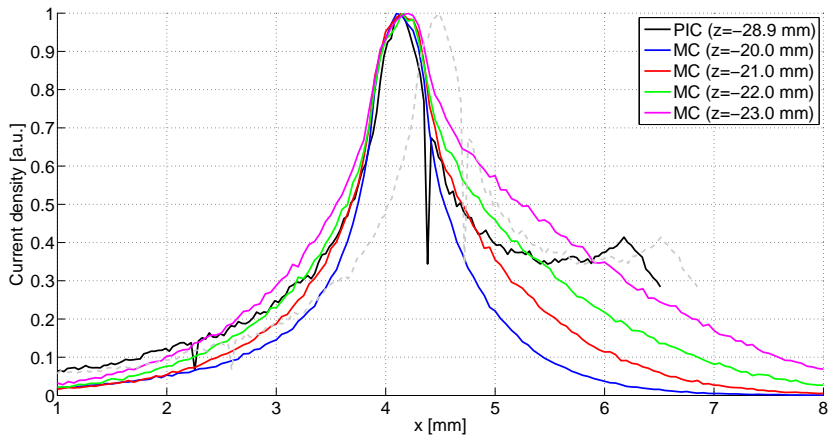}
\caption{Profiles of current density along $x$ direction for PIC (black) and MC (colored) simulations. The PIC profile was shifted by 0.34 mm to facilitate comparison with MC results, original PIC profile shown in grey.}
\label{PIC_vs_MC_cur}
\end{center}
\end{figure}
Current density profiles obtained by both codes are shown in Fig. \ref{PIC_vs_MC_cur} (particles were injected at $x$ = 0.0). The position of the peak in PIC results was shifted by 0.34 mm with respect to MC profiles, so in order to facilitate comparison of the two methods, it was shifted respectively. Since the fit of the $T_i$ is done with respect to the peak maximum, this does not influence further analysis. It can be seen, that PIC profiles best match MC profile at $z$ = $-22.0$ (green line) but the MC profile is not strongly dependent on $z$ (note that z=0 represents slit position and in PIC simulation the electrodes are located between $z$ = -4.0 and -30.0 mm). Although the profiles are not completely identical, one has to keep in mind that for the analysis only the rising (left) slope is being used, and that the current is collected by collectors typically $\sim$0.9 mm wide, for which the agreement is sufficient. Since the computational demands of the PIC simulations does not allow for large number of runs, which are needed for calibration of the results,  In the further analysis we will rely on the MC code with shortened cavity length $a$ by 3.0 mm. The MC code simulates an infinitely thin slit plate with rectangular slit. Originally, the slit cross-section 1.0$\times$0.1 mm$^2$ was used, however this simplification was later refined to account for knife-edge profile and Larmor losses at slit edges. An effective slit length $L_{eff}$ was calculated using slit length in the middle of slit plate, from which ion Larmor radius $r_{Li}$ was substracted on each side. For COMPASS parameters, this resulted in reduction from 1.0 to 0.5 mm.

\begin{figure}[htbp]
\begin{center}
\includegraphics[scale=0.33]{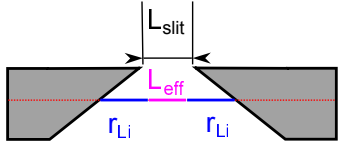}
\caption{Calculation of the effective slit length $L_{eff}$.}
\label{eff_slit}
\end{center}
\end{figure}

Using the MC code, the current density profile was mapped onto an array of synthetic collectors. Depending on the collector current distribution, two techniques can be used to fit the $T_i$. If the rising edge of the current profile is distributed on more than 2 collectors, $T_i$ is obtained using an exponential fit to the collector currents, where collector positions are translated to parallel velocities using equations \ref{def} and \ref{eq-acc}. This technique allows also to obtain an estimation of error of the fit. If the rising edge covers only two collectors, a simple 2-point interpolation is used:
\begin{equation}
%T_i = \frac{log(I_c(7)/I_c(6))} {2 (v_{||}^2(7) - v_{||}^2(6))}.
T_i = \frac{m_i E L}{2 e B_{loc}}  \left (\frac{1}{\Delta^2_x(2)} - \frac{1}{\Delta^2_x(1)}  \right ) / log \left (\frac{I_{col}(1)}{I_{col}(2)} \right),
\label{fit}
\end{equation}  
where indexes $1$ and $2$ denote the two collectors respectively.

\begin{figure}[htbp]
\begin{center}
\includegraphics[scale=0.33]{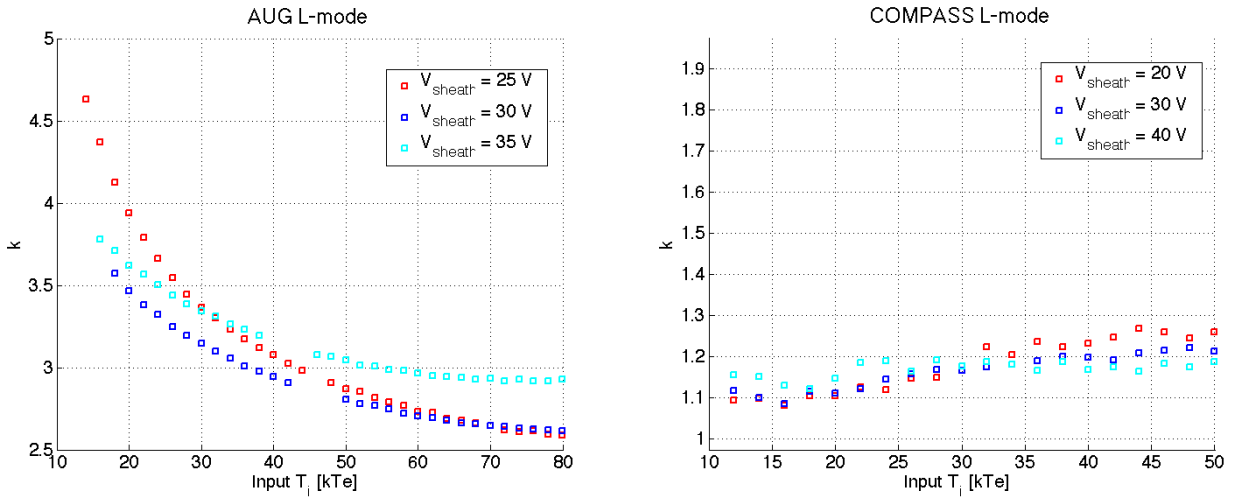}
\caption{Scaling of the $k$ parameter:  for AUG L-mode conditions (left) obtained by 2-point interpolation method ($E_y$ = 33 kV/m) and for COMPASS L-mode conditions (right) obtained by exponential fit ($E_y$ = 50 kV/m) for different values of plasma potential ($V_{sheath}$). }
\label{k-lmode}
\end{center}
\end{figure}

 The obtained ion temperature was compared with the temperature of injected ions. The systematic difference between these two values was characterized by coefficient $k$
\begin{equation}
              T_{i_{real}} = k T_{i_{simulated}},
\end{equation}
which in principle can depend on a large number of parameters -  analyzer's dimensions, electric field applied between the electrodes $E_y$, tokamak magnetic field B, potential between the electrodes along the ion beam $V_{middle}$ and ion temperature itself. The quality of the fit also depends on the size and spacing of the collectors and amount of parasitic noise in acquired signals. In the design consideration some of the parameters (such as the $B$ field and cavity size) are determined by constraints specific to each machine, some (such as the  $E_y$ field) can be adjusted to improved the quality of acquired signals. The aim of this study is not to determine an universal scaling of the $k$ parameter, which could be used for any tokamak but rather than that present scaling for two machines, where the specific probe head was utilized - COMPASS and AUG (see Fig. \ref{k-lmode}). Since the probe head is not capable of measuring the local plasma potential, it remains a free parameter in the scaling. However, for L-mode studies, it has been observed \cite{adamek-lmode} in AUG that the fluctuations of plasma potential are relatively small, typically around +30V. The obtained scaling of $k$ parameter does not show a substantial variation of the $k$ parameter with plasma potential, which is due to condition $|V_{mid}| > |V_{plasma}|$. The scaling of $k$ parameter for AUG was obtained using the 2-point interpolation technique using voltage difference 200 V between the electrodes ($E_y$ = 33 kV/m), for COMPASS the exponential fitting has been used with voltage difference 300 V ($E_y$ = 50 kV/m).

\subsection{Electron collector current}
The first experiment revealed parasitic electron current on all collectors, which was interfering with the measurements.  The origin of the electrons inside the cavity was not immediately evident, however an effective mean on reducing the current was found out in experiment. The voltages on electrodes were arranged so that  $V_{mid} < V_{slit}$. This setup however enhanced acceleration of ions and consequently reduced the dispersion due to applied electric field. An alternative solution was to introduce a protective grid in front of the collector array. When this grid was biased so that $V_{grid} <  V_{slit}$, the electron current was successfully suppressed. This indicated that the electrons are secondaries originating from the back side of the slit plate and thus repelled by a potential lower than $V_{slit}$. Such secondary electrons would however have to be created by impact of either ions or electrons on the back side of the slit, which was in contradiction with a simple model of the analyzer's behavior. The particle-in-cell simulations, which were originally used to assess the space charge effects inside the cavity, revealed that due to strong E field inside the slit volume, the ion trajectories are deformed so that some ions are reflected towards the back side of the slit plate (example trajectories are shown in Fig. \ref{trajectories} right).

\section{Design of the analyzer}

\begin{figure}[htbp]
\begin{center}
\includegraphics[scale=0.33]{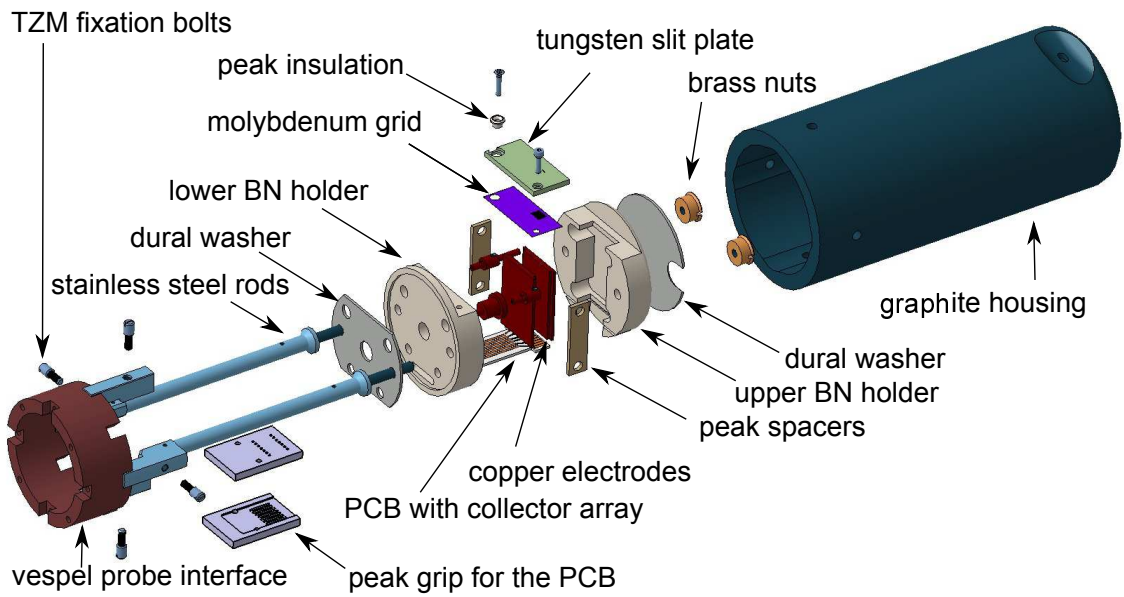}
\caption{Schematic view of the ExB analyzer probe head.}
\label{rozstrel}
\end{center}
\end{figure}

 The analyzer designed for AUG and COMPASS tokamak (as shown in Fig. \ref{rozstrel}) is contained within a 5 mm thick carbon cylindrical housing (60 mm in diameter, 150 mm length), which is attached to the internal structure by four TZM fixation bolts. The cavity entrance is located 15 mm bellow the probe head ending, where the housing thickness is reduced to 3 mm around an orifice of approx. 12 mm$^2$.  Since the dimensions of the orifice (approx. 5$\times$3 mm) is larger than ion Larmor radius ($r_L <$ 1 mm), we assume that the ion flux arriving into the slit is not attenuated by the housing. We also use the orifice area as an effective collecting area for ions falling onto the slit plate, neglecting possible finite Larmor effects.

The entrance slit is machined into a 2 mm thick  tungsten plate following the design described in section \ref{section-slit} using the spark erosion technique. First measurements were performed with a 1 mm long and 100 $\mu$m wide slit with edge opening at  20$^\circ$.  The slit is oriented with its longer side along the collector array. 

The internal cavity is made of Boron Nitride parts with two copper planar electrodes. The electrodes covered 30 out of the 36 mm of total distance from slit to the collector array. The electrode separation  is 6 mm, which is more than the expecred ion Larmor radii and allows to create sufficient electric fields ($\sim$ 50 kV/m) using available power sources without an elevated risk of arcing (up to 400V).

\begin{figure}[htbp]
\begin{center}
\includegraphics[scale=0.5]{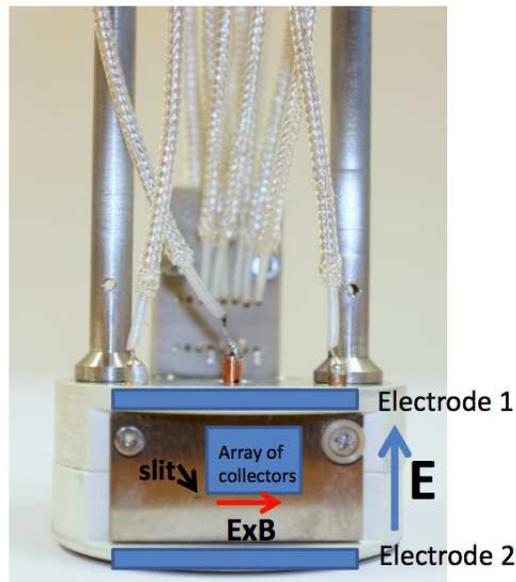}
\caption{Front view of the analyzer with the tungsten slit plate in front.}
\label{photo}
\end{center}
\end{figure}

The collector array is printed on a FR4 PCB (made by Pragoboard company), which has significant advantages over other approaches, such as the sandwich design\cite{renaud-sandwich}. Risk of short-circuits between the segments is minimal, it occupies less space and in principle segments with varying size are easy to produce. For the first experiment we used an array of 12 collectors, each 0.9 mm wide with 0.3 separation.  The number of segments is limited by the number of pins in the manipulator interface (22 in AUG, 18 in COMPASS) and number of available channels in fast data acquisition system (8 during the first experiment). The upper part of the analyzer is connected to the manipulator interface by two stainless steel rods (see Fig. \ref{photo}). 

The signals from collectors are carried by standard vacuum-compatible Huber-Suhner cables to the manipulator interface, which consists of a matrix of SNB connectors. The cables run through the manipulator towards the current pre-amplifier (total length of cables from collectors to the amplifier is $\sim$4 meters).

\subsection{Voltage setup}
The following arrangement of voltages was used in the first experiments, which were focused on measurements of turbulence in SOL. The slit plate was biased to -160 V and the grid in front of the collector array to -180 V. The top electrode was left at 0 V, while the bottom electrode was set to -200V. Assuming that the plasma potential is positive in SOL [REF Adamek], the potential inside the cavity is $< V_{plasma}$ and so ions can freely propagate. In order to determine $\Delta_x$, the knowledge of precise alignment of local B field with the analyzer is needed in order to determine the position at which field lines passing through the slit intersect the collector array. Such precise alignment ($<1^\circ$) cannot be measured a priori, since the orientation of field lines depend on local value of safety factor $q$ but  has to be determined in the experiment for a given location of the probe. In order to do so, a short synchronouse voltage pulse was introduced on all components, which dropped the voltage to 0V. During the pulse, electrons can propagate inside the cavity along the field line and mark the collector, which is aligned with the slit.

\section{Summary}
An instrumental study of the E$\times$B analyzer for COMPASS and AUG tokamaks has been performed by means of MC and PIC simulations. The geometry of the slit with knife-edge opening at 20 degrees has been found convenient in terms of its transmission and power handling. A series of MC simulations was used to identify combinations of applied $E$ field and bias voltage, which are suitable for  $T_i$ measurements in the conditions expected in the SOL. PIC simulations we performed in order to investigate the possibility of a space charge build up in the entrance slit. The analyzer was successfully constructed and tested on AUG to obtain temperature measurements at the timescale of blobs. 
\medskip 
\section{Acknowledgements}
This work was partly supported by MSMT Project $\#$LM2011021 and by EFDA Fellowship programme.

%\footnotesize

\end{document}